\documentclass[aps,groupedaddress]{revtex4}

\usepackage{graphicx}
\begin{document}

\author{F.\ Marty Ytreberg}
\author{Daniel M.\ Zuckerman}
\affiliation{Center for Computational Biology and Bioinformatics,
  University of Pittsburgh, 200 Lothrop St., Pittsburgh, PA 15261}
\date{\today}
\title{Efficient use of non-equilibrium measurement to estimate 
  free energy differences for molecular systems}
\date{\today}

\begin{abstract}
  A promising method for calculating free energy
  differences $\Delta F$ is to generate
  non-equilibrium data via ``fast-growth'' simulations
  or experiments -- and then use Jarzynski's equality.
  However, a difficulty with using Jarzynski's equality is that
  $\Delta F$ estimates converge very slowly and unreliably due to
  the nonlinear nature of the calculation -- thus requiring large, costly
  data sets. Here, we present new analyses of non-equilibrium data
  from various simulated molecular systems exploiting statistical
  properties of Jarzynski's equality. Using a fully automated
  procedure, with no user-input parameters, our results suggest that good
  estimates of $\Delta F$ can be obtained using 6-15 fold less
  data than was previously possible. 
  Systematizing and extending previous work \cite{zuckerman-cpl},
  the new results exploit
  the systematic behavior of bias due to finite sample size.
  A key innovation is better use of the more statistically
  reliable information available from the raw data.
\end{abstract}


\maketitle

\section{Introduction}
The calculation of free energy differences, $\Delta F$
plays an essential role in
many fields of physics, chemistry and biology
\cite{zuckerman-cpl,bustamante-sci,zuckerman-prl,shirts-prl,pearlman,kong,
shirts-jcp,bustamante-pnas,hummer-pull,bash-sci,woods-jpcB,mccammon,
shobana,isralewitz,karplus-cpl,karplus-jpcB,schulten-2003}.
Examples include determination of the solubility of
small molecules, and binding affinities of ligands to proteins.
Rapid and reliable estimates of $\Delta F$ would be particularly
valuable to structure-based drug design, where current approaches to
virtual screening of
candidate compounds rely primarily on ad-hoc methods \cite{burgers,bajorath}.
Free energy estimates are also critical for protein engineering
\cite{mayo,degrado}.

The focus of this report is non-equilibrium ``fast-growth'' free energy
calculations \cite{zuckerman-cpl}.
These methods hold promise -- yet to be fully realized --
for very rapid estimation of $\Delta F$.
The central idea behind the non-equilibrium methods
is to calculate the {\it irreversible}
work during a very rapid (thus non-equilibrium) switch
between the two systems or states of interest.
Multiple switches are done, and the resulting set of work values
can be used to estimate $\Delta F$ using
Jarzynski's equality (detailed in Sec.\ \ref{sec-fastgrowth})
\cite{jarzynski-both}.

Somewhat surprisingly, non-equilibrium $\Delta F$ calculations
are critical for analyzing single-molecule pulling {\it experiments}
\cite{bustamante-sci,hummer-pull}. In essence, these experiments
generate non-equilibrium work values, as pointed out by Hummer and
Szabo, so the only way to estimate the free energy profile
is to use Jarzynski's equality \cite{hummer-pull,schurr}.
The methods that we develop in this report should be equally useful
for analyzing such experiments.

It has been accepted for some time that there are three sources of error
\cite{zuckerman-jstat}
for non-equilibrium $\Delta F$ calculation:
(i) inaccuracy of the
force field \cite{grossfield-jacs},
(ii) inadequate sampling of the configurational space
\cite{lu-i,lu-ii,oostenbrink,jarzynski-targeted,mordasini}, and
(iii) bias due to finite sample size
\cite{zuckerman-cpl,wood,jarzynski-master,miller,hu-ms,schon}.
Error in free energy calculations have been of long-standing interest,
e.g.\ Refs.\ \cite{brunger,pearlman-lag,dinola,pearlman-1994,edholm}

The present study addresses only source (iii), and
attempts to determine the most efficient use of
fast-growth work values. In other words, given a (finite)
set of work values generated by simulation or experiment,
what is the best estimate for $\Delta F$?
We do not here attempt to prescribe the best method for generating
non-equilibrium work values.

We proceed by first introducing two
new block averaging techniques, based on the original proposal
by Wood {\it et al}.\ \cite{wood-block}. Block
averaging provides well-behaved, but biased, $\Delta F$ estimates.
We then discuss two distinct schemes for extrapolating
to the ``infinite data limit''. Our work systematizes and extends
previous work by Zuckerman and Woolf \cite{zuckerman-cpl},
who originally proposed the use of block averages for extrapolation.

Methods to lessen the effect of bias due to finite sample size
have been proposed for the case when switching
between systems is performed in both directions \cite{lu-2001,lu-2003,bennett},
and for the simplified case in which the non-equilibrium
work values follow a quasi-Gaussian distribution
\cite{bustamante-pnas,amadei-moments}.
Hummer also considered errors in non-equilibrium $\Delta F$
calculations \cite{hummer}. To our knowledge, however, other
workers have not addresses uni-directional switching in
highly non-Gaussian systems.

The techniques outlined in the following sections offer
rapid estimates for $\Delta F$ for the systems
we studied -- namely, the chemical potential for a
Lennard-Jones fluid,
``growing'' a chloride ion in water, methanol $\rightarrow$
ethane in water, and palmitic $\rightarrow$ stearic acid in water.
Work values for these systems follow highly non-Guassian distributions.
We compare our extrapolated results to $\Delta F$ obtained
by using Jarzynski's equality,
finding a 6-15 fold decrease in the
the number of work values needed to estimate $\Delta F$ for the
test systems considered here.

\section{\label{sec-fastgrowth} Fast-Growth}
Fast-growth techniques have been described in detail elsewhere
\cite{zuckerman-cpl,jarzynski-both,hummer}, so
we will simply outline the method.
Consider two systems defined by potential energy functions
$U_0$ and $U_1$. To calculate the free energy difference $\Delta F$
between these two systems, one must simply ``switch'' the system from 
$U_0$ to $U_1$. This is readily accomplished by defining a switching
parameter $\lambda$ such that
\begin{eqnarray}
  U_{\lambda} ({\bf x})=U_0({\bf x})+
  \lambda \Bigl[U_1({\bf x})-U_0({\bf x})\Bigr],
  \label{eq-U_lambda}
\end{eqnarray}
where {\bf x} is a set of configurational coordinates, and
$U_{\lambda}({\bf x})$ describes the ``hybrid'' potential energy
function for all values of $\lambda$ from 0 to 1.
We note that nonlinear scaling with $\lambda$ is also
possible \cite{kong,pitera,karplus-jpcB,karplus-cpl,shirts-jcp}
resulting in hybrid potentials
differing from Eq.\ (\ref{eq-U_lambda}).
Our approach here also applies, in principle,
to other such choices.
Essentially, the idea behind fast-growth methods
is to perform rapid switches from
$\lambda=0 \rightarrow 1$, where each
switch is generated starting from coordinates
drawn from the equilibrium ensemble for $\lambda=0$.
During each switch, the irreversible work is accumulated,
generating a single work value.
Multiple switches are done to generate a distribution
of these work values $\rho(W)$.

\subsection{\label{sec-estimators} Simple Estimators}
It has been appreciated for some time that the average work
obtained over many such switches provides a rigorous upper bound for 
the free energy difference,
\begin{eqnarray}
  \Delta F \leq \bigl< W \bigl>_0,
  \label{eq-Wave}
\end{eqnarray}
where the $\bigl<...\bigl>_0$ represents an average over
many switches starting from the equilibrium ensemble for
$\lambda=0$ and ending at $\lambda=1$. Equality occurs only
in the limit of infinitely slow switches.
Further, if the distribution of work values
$\rho(W)$ is Gaussian (this occurs if the
system remains in equilibrium during the switch, but
also may occur in certain far from equilibrium
situations), then the 
high temperature expansion of Zwanzig \cite{zwanzig}
gives
\begin{eqnarray}
  \Delta F = \bigl< W \bigl>_0-\frac{1}{2}\beta\sigma_W^2,
  \label{eq-gauss}
\end{eqnarray}
where $\sigma_W$ is the standard deviation of $\rho(W)$,
and $\beta=1/k_BT$ where $T$ is the temperature
of the system and $k_B$ is the Boltzmann constant.

However, for the fast-growth work values under consideration here,
the distribution of work values can be very broad and
non-Gaussian. Thus
Eqs.\  (\ref{eq-Wave}) and (\ref{eq-gauss})
will not provide reasonable estimates of $\Delta F$; see Fig.\
\ref{fig-hist}.
It is possible to use higher order
moments to estimate $\Delta F$ (see for example Refs.\ 
\cite{zwanzig,bustamante-pnas,hummer-moments,amadei-moments}). These
estimators are most useful in the near-equilibrium
regime.

\subsection{\label{sec-jarz} Jarzynski Equality}
Due to recent work by Jarzynski
\cite{jarzynski-both,hendrix,crooks-pre}, it
is possible to estimate $\Delta F$ using
these fast-growth $W$ values via
\begin{eqnarray}
  {\rm e}^{-\beta\Delta F}=
  \Bigl<{\rm e}^{-\beta W}\Bigr>_0,
  \label{eq-jarzequal}
\end{eqnarray}
This remarkable relationship is valid for {\it arbitrary}
switching speed, implying that one can perform switches as
rapidly as desired and still obtain valid estimates of
$\Delta F$. The Jarzynski equality thus provides an
estimate for $\Delta F$ for a set of $N$ work values given by
\begin{eqnarray}
  \Delta F_{Jarz}=-\frac{1}{\beta}\ln\Biggl[
    \frac{1}{N}\sum_{i=1}^N {\rm e}^{-\beta W_i}
    \Biggr],
  \label{eq-jarz}
\end{eqnarray}

\begin{figure}
\includegraphics[scale=0.5]{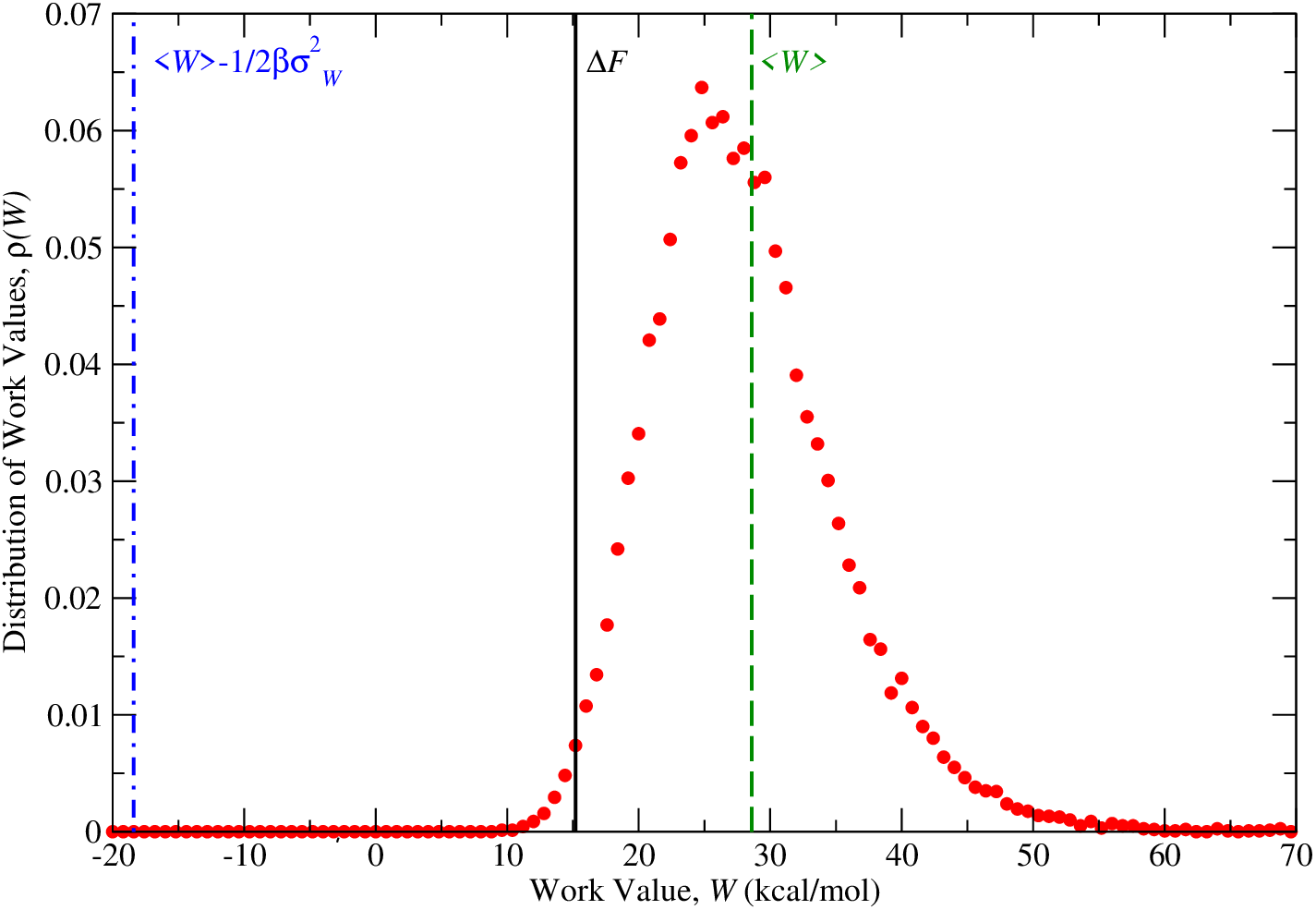}
  \caption{\label{fig-hist}
    Distribution of work values for PAL2STE test system
    (palmitic $\rightarrow$ stearic acid mutation in water,
    described in Sec.\ \ref{sec-systems}). Also included
    in this plot are the estimators given by Eqs. (\ref{eq-Wave})
    and (\ref{eq-gauss}) shown by the blue dot-dash and green dashed line
    respectively.
    The solid black line shows the $\Delta F$ estimate obtained by using
    Jarzynski's equality in Eq.\ (\ref{eq-jarz}) for all available data.
  }
\end{figure}

The $\Delta F$ estimates given by Eq.\ (\ref{eq-jarz}), however,
are very sensitive to the distribution of
work values $\rho(W)$
\cite{zuckerman-cpl,wood,jarzynski-master}.
If the width of the work distribution large,
i.e.\ $\sigma_W \gg k_B T$
(this implies a very rapid switch and/or a complex system),
then often thousands, or even
tens of thousands of work values are needed to reliably estimate
$\Delta F$. An example of this can be seen in Fig.\ \ref{fig-hist}
where a histogram of work values is shown for PAL2STE
(described in Sec.\ \ref{sec-systems}). The value of $\Delta F_{Jarz}$
given by the Jarzynski equality, as well as
estimators $\bigl< W \bigl>_0$ and
$\bigl< W \bigl>_0-\frac{1}{2}\beta\sigma_W^2$,
are shown on this plot. This graphically demonstrates why
Eqs.\ (\ref{eq-Wave}) and (\ref{eq-gauss}) are often poor estimates
of the free energy for fast-growth work values.

If the switch is performed instantaneously, then
Eq.\ (\ref{eq-jarz}) becomes
\begin{eqnarray}
  {\rm e}^{-\beta \Delta F}=
  \biggl<{\rm e}^{-\beta \bigl[U_1({\bf x})-U_0({\bf x})\bigr]}\biggr>_0,
  \label{eq-fep}
\end{eqnarray}
often called single-stage free energy perturbation
\cite{zwanzig,kirkwood}.
In this limit, the system is not allowed to relax at any 
intermediate values
of $\lambda$. Instead $U_1({\bf x})$ is simply evaluated at
values of ${\bf x}$ drawn from the equilibrium ensemble for
$\lambda=0$. The advantage of this method is that data
can be generated very quickly. However, in practice,
unless there is sufficient overlap between the states
described by $U_0$ and $U_1$, the estimate of $\Delta F$
will be biased, often by many $k_B T$
\cite{liu,swanson}.
The problem of attaining overlap of states can be improved
by drawing from the equilibrium ensemble for
an unphysical ``soft-core'' state (such as for $\lambda=0.5$)
\cite{oostenbrink,pitera}.

Recent work by Hendrix and Jarzynski
\cite{hendrix} showed that essentially the only determining factor in 
the accurate calculation of $\Delta F$ was physical CPU time
spent during the calculation. So, doing many rapid switches 
has no advantage over doing fewer slower switches.
This conclusion is based upon using Eq.\ (\ref{eq-jarz})
for all $\Delta F$ estimates.

This manuscript describes methods that exploit
statistical properties of Jarzynski's equality, allowing us
to do use work values from very rapid switches and
obtain $\Delta F$ estimates with 6-15 fold less work
values than using Eq.\ (\ref{eq-jarz}).

\section{\label{sec-slowgrowth} Other Methods}
To calculate reliable $\Delta F$ estimates using less work values,
one can generate a narrower $\rho(W)$,
i.e.\ perform the switching process more slowly. However, slower
switching speed means that more computational time will be spent
to generate each work value -- offsetting some of the
advantage gained by doing rapid switches.

If the switch is performed so slowly that the system
remains near equilibrium during the switch, then the width
of the distribution will be very small ($\sigma_W < k_B T$),
and thus only a few work values are required for accurate
estimation of $\Delta F$ \cite{hermans,hu-ms}.
This slow-growth method is, in principle,
equivalent to thermodynamic integration \cite{leach-book} where
$\Delta F$ is calculated by
allowing the system to reach equilibrium for each value
of $\lambda$. Then $\Delta F$ is found using
\begin{eqnarray}
  \Delta F=
  \int_0^1 d \lambda \Biggl<
  \frac{\partial U_{\lambda}({\bf x})}{\partial \lambda} \Biggr>_{\lambda}.
  \label{eq-ti}
\end{eqnarray}
Thermodynamic integration and slow-growth
can provide very accurate $\Delta F$
calculations, but are also computationally expensive
\cite{karplus-jpcB,mordasini,shirts-jcp,lybrand}.

As previously mentioned, the equilibrium ensemble, when using the
Eq.\ (\ref{eq-jarz}), is generated for $\lambda=0$. Then $\rho(W)$
is generated by doing switches from $\lambda=0 \rightarrow 1$
(forward switches) with configurations drawn from
the $\lambda=0$ ensemble. It is also possible
to generate another equilibrium ensemble for $\lambda=1$ and then
perform {\it reverse} switches from $\lambda=1 \rightarrow 0$.
It has been shown that, if one combines the use of the
forward and reverse work values, convergence is much more
rapid then doing just forward switches
\cite{bennett,lu-2001,lu-2003,frenkel-book}.
It has been recently
demonstrated that most efficient use of forward and
reverse work values is for Bennett's method
\cite{shirts-prl,lu-2001,lu-2003}.

There is, however, a distinct advantage to using
Jarzynski's estimates with only forward switches,
when one considers the eventual
goal of predicting relative binding affinities for
application in drug design. In this situation, if
using Jarzynski estimates, one need only generate
a single high-quality equilibrium ensemble
for a particular ligand-receptor or reference
complex. Then one can determine relative binding affinities
for other ligands without generating another equilibrium
ensemble -- a significant decrease in computational expense.

\section{\label{sec-systems} Test Systems}
To show the generality of the methods proposed in this study
we consider four test systems with varying molecular complexity:
a chemical potential calculation for a Lennard-Jones fluid, 
``growing'' a chloride ion in water,
methanol $\rightarrow$ ethane in water,
and stearic $\rightarrow$ palmitic acid in water.

The last two systems are alchemical mutations of fully 
solvated molecules (see Refs.\ \cite{zuckerman-cpl,nanda}
for simulation details), and the first system is a chemical potential
calculation done by the particle insertion method (see Ref.\
\cite{hendrix} for details). All three of these data
sets were generated previous to this study \cite{thanks}.

The growing chloride system was studied using
TINKER version 4.1
\cite{ponder}, with the simulation conditions 
chosen to closely match those of
Lybrand {\it et al}.\ in Ref.\ \cite{lybrand}.
Stochastic dynamics
simulations were carried out in the canonical
ensemble (constant $N,V,T$) in a cubic box of edge
length 18.6216 \AA. The temperature was held
at 300 K by a Berendsen thermostat with a time constant
of 0.1 \cite{berendsen}.
The chloride ion was modeled with Lennard-Jones
parameters $\sigma=4.4463$ \AA\, and
$\epsilon=0.1070$ kcal/mol, and was solvated by 214 SPC water
molecules. Ewald summation approximated charge interactions and
RATTLE was used to hold the water molecules rigid \cite{andersen}.
For this test system, the Lennard-Jones
``size'' was increased by 1.0 \AA, from $\sigma=4.4463$ \AA\,
at $\lambda=0$ to $\sigma=5.4463$ \AA\, at $\lambda=1$.

To obtain fast-growth work values, a time step of 1.0 fs was
used. The system was equilibrated
for at least 10 ps, after which starting
configurations for each fast-growth trajectory
were generated every 100 time steps.

Below we list the notation used to refer to each data set. Also
included are statistical features of the data sets -- the total
number of work values ($N_{tot}$), the
mean work ($\bigl< W \bigr>$) and the standard deviation ($\sigma_W$).
These data sets are all considered difficult to use for $\Delta F$
calculations owing to the facts that $\sigma_W \gg k_BT$
and $\bigl< W \bigr>-\Delta F > 10 k_BT$;
see Eqs.\ (\ref{eq-Wave}) and (\ref{eq-gauss}) and Fig.\ \ref{fig-hist}.
\begin{list}{}{}
\item LJ -- Chemical potential calculation for a Lennard-Jones fluid
in 1 $\lambda$-step \cite{hendrix}.
This corresponds to instantaneous switching or
free energy perturbation, as described in sec.\ \ref{sec-fastgrowth}. 
$N_{tot}=100,000$, $\bigl<W\bigr>=305.1$ $k_BT$ and $\sigma_W=83.5$ $k_BT$.
Using all work values, Eq.\ (\ref{eq-jarz}) gives a best estimate
$\Delta F_{best}=0.7$ $k_BT$.
\item GROWCL -- Grow chloride by 1.0 \AA\, in
10 $\lambda$-steps with 1 relaxation step at each value of $\lambda$.
$N_{tot}=40,000$, $\bigl<W\bigr>=40.1$ kcal/mol and $\sigma_W=8.6$ kcal/mol.
Using all work values, Eq.\ (\ref{eq-jarz}) gives $\Delta F_{best}=18.4$ kcal/mol.
\item METH2ETH -- Methanol to ethane mutation data using
200 $\lambda$-steps
with 1 dynamic relaxation step at each value of $\lambda$ \cite{zuckerman-cpl}.
$N_{tot}=9,600$, $\bigl<W\bigr>=37.0$
kcal/mol and $\sigma_W=12.3$ kcal/mol. Using all
work values, Eq.\ (\ref{eq-jarz}) gives $\Delta F_{best}=7.4$ kcal/mol.
\item PAL2STE -- Palmitic to Stearic acid mutation data using
55 $\lambda$-steps
with 10 relaxation steps at each value of $\lambda$ \cite{nanda}. 
$N_{tot}=20,000$, $\bigl<W\bigr>=28.6$ kcal/mol and $\sigma_W=7.5$
kcal/mol. Using all
work values, Eq.\ (\ref{eq-jarz}) gives $\Delta F_{best}=15.2$ kcal/mol.
\end{list}

Since the goal is to determine the best analysis
for a given set of work values, we assume that the true $\Delta F$
is given by Eq.\ (\ref{eq-jarz}) using all available work values
(i.e.\ $\Delta F_{best}$ above).
Determining whether the distribution of work values $\rho(W)$
used in this paper 
are complete and representative is beyond the scope of this report.

\section{\label{sec-block} Block Averaging}
\begin{figure}
\includegraphics[scale=0.5]{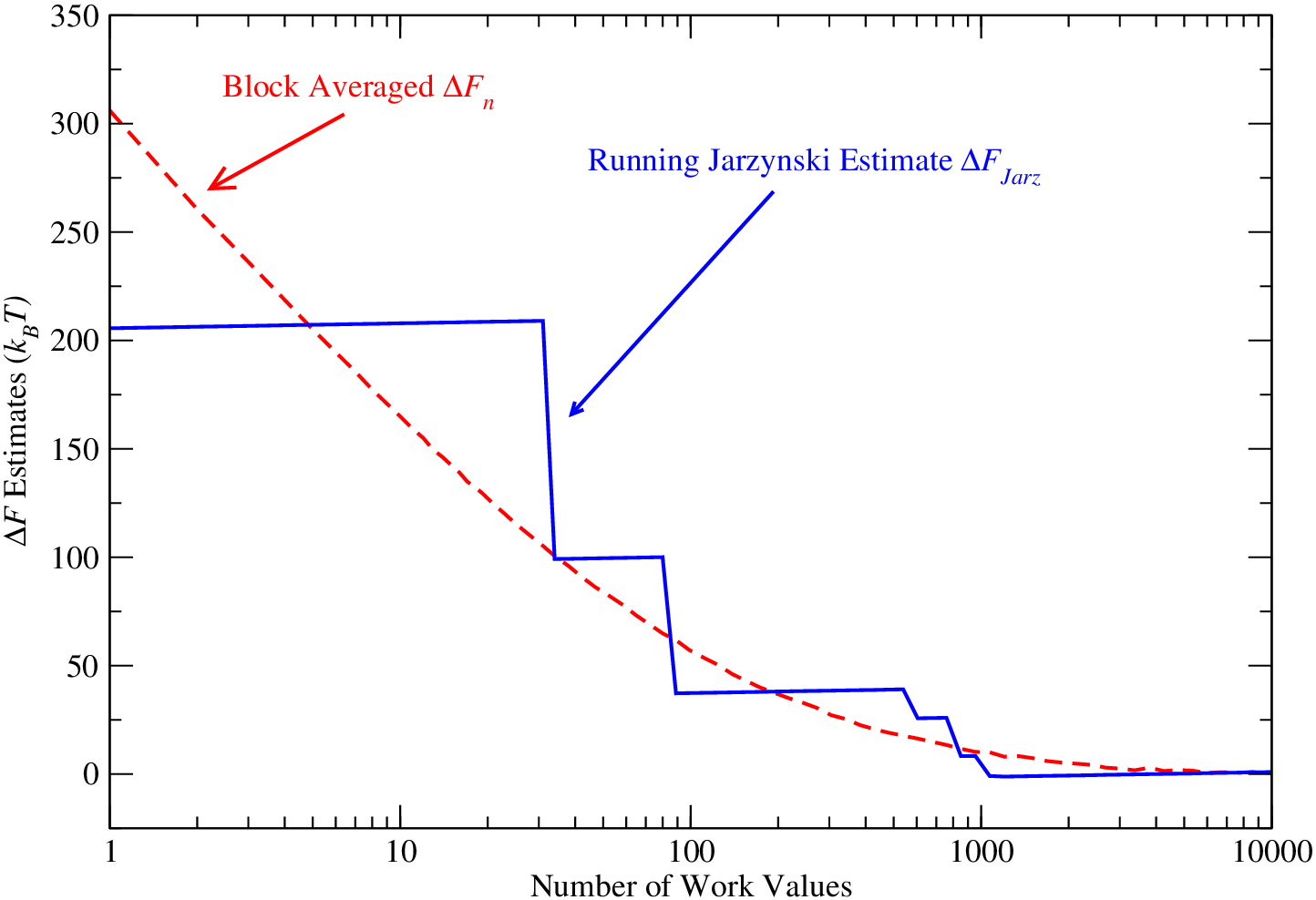}
  \caption{\label{fig-block}
    The running Jarzynski estimate, given by Eq.\ (\ref{eq-jarz}),
    as a function of the number of work values
    used in the estimate, $N$ is shown as a solid blue line.
    The dashed red line shows the sub-sampled block averaged free
    energy estimate given by Eq.\ (\ref{eq-block}), 
    plotted as a function of the number of work values
    in each block, $n$. Data used for these estimate
    were obtained from the LJ test system (chemical potential
    for a Lennard-Jones fluid). The Jarzynski estimate displays
    erratic convergence behavior, while the block averaged free
    energy estimate displays a smooth monotonically
    decreasing estimate
  }
\end{figure}

The motivation for using block averages can be seen in Fig.\ 
\ref{fig-block} (see also Refs.\ \cite{zuckerman-prl,wood-block}).
The solid blue line is a running estimate for $\Delta F$
obtained by using Eq.\ (\ref{eq-jarz}) and the
dashed red line is obtained by block averaging. Both curves are obtained
using the LJ test system. The running Jarzynski estimate
exhibits very poor convergence behavior, making it very difficult to
establish when a reliable estimate of $\Delta F$ has
been obtained.
The block averaged free energy, however,
displays a smooth monotonically decreasing
$\Delta F$ estimate, which approaches the true $\Delta F$.

Each block averaged free energy ($\Delta F_n$) 
data point was obtained from a set of $N$ work values
($W_1,W_2,...,W_N$) using the following scheme
\cite{wood-block}:
\begin{enumerate}
\item Draw $n$ work values at random from the set, generating a subset
  ($W_1,W_2,...,W_n$).
  This is now the $j^{\rm th}$ block of work values.
\item Use Jarzynski's equality, Eq.\ (\ref{eq-jarz}) to obtain a free energy
  estimate $F_j$ for this block
  \begin{eqnarray}
    F_j=-\frac{1}{\beta}\ln\Biggl(\sum_{i \in \,{\rm block}\, j}
    {\rm e}^{-\beta W_i}\Biggr).
  \end{eqnarray}
\item Repeat steps 1 and 2 until you have $m$ blocks, each containing
  $n$ values. Now the average ($\Delta F_n$)
  and standard deviation ($\sigma_n$) can be calculated using
  \begin{eqnarray}
    \label{eq-block}
    \Delta F_n=\frac{1}{m}\sum_{j=1}^m F_j =
    \frac{1}{m}\sum_{j=1}^m\Biggl[ -\frac{1}{\beta}\ln\Biggl(\sum_{i \in j}
      {\rm e}^{-\beta W_i}\Biggr) \Biggr],
  \end{eqnarray}
  \begin{eqnarray}
    \sigma_n^2=\frac{n}{N}\sum_{j=1}^m (F_j-\Delta F_n)^2.
    \label{eq-blockSn}
  \end{eqnarray}
\end{enumerate}
This process is carried out for every possible value of $n$
(i.e.\ $n=1,2,3,...,N$).

In previous work \cite{zuckerman-cpl,wood-block},
$m=N/n$ was chosen, i.e.\ $m$ is the number of
blocks of size $n$ from a data set of size $N$. The
weakness of this choice is that a reshuffling of
the data set gives a new (generally different)
set of $\Delta F_n$ values.
To avoid this weakness we choose $m$ large enough that
the resulting $\Delta F_n$ values do not depend upon
the value of $m$. This is typically accomplished with
$m \sim 100\times N/m$.

Since there are two distinct ways of randomly drawing from
a data set (i.e.\ implementing the first step above),
we introduce two new block averaging schemes.
The first is to draw work values
from ($W_1,W_2,...,W_N$) at random {\it with replacement} --
i.e.\ it is possible to draw a particular work value more than once.
We call this a bootstrapped $\Delta F_n$ \cite{bootstrap-book}.
The second is to draw from ($W_1,W_2,...,W_N$)
at random {\it without replacement}. We call
this a sub-sampled $\Delta F_n$ \cite{subsample-book}.

The difference between the bootstrapped and sub-sampled methods can
be illustrated by considering a data set of $N$ work values where $N-1$ 
values are large and one value is very small. Due to the highly nonlinear
nature of the Jarzynski equality, the single small work value will dominate
Eq.\ (\ref{eq-block}). Suppose one calculates $\Delta F_n$
for $n=N$ using both of these methods. The sub-sampled method will only 
have one $\Delta F_N$ estimate since reshuffling the work values
has no effect when $n=N$. However, the bootstrapped method calculates
a $\Delta F_N$ value that is larger than the sub-sampled $\Delta F_N$
due to the fact that it will draw the
small work value only a fraction of the time.
A generalization of this argument shows that
the bootstrapped $\Delta F_n$ will exceed the sub-sampled
$\Delta F_n$ for every value of $n$.  

\section{Extrapolation Methods}
\begin{figure}
  \includegraphics[scale=0.5]{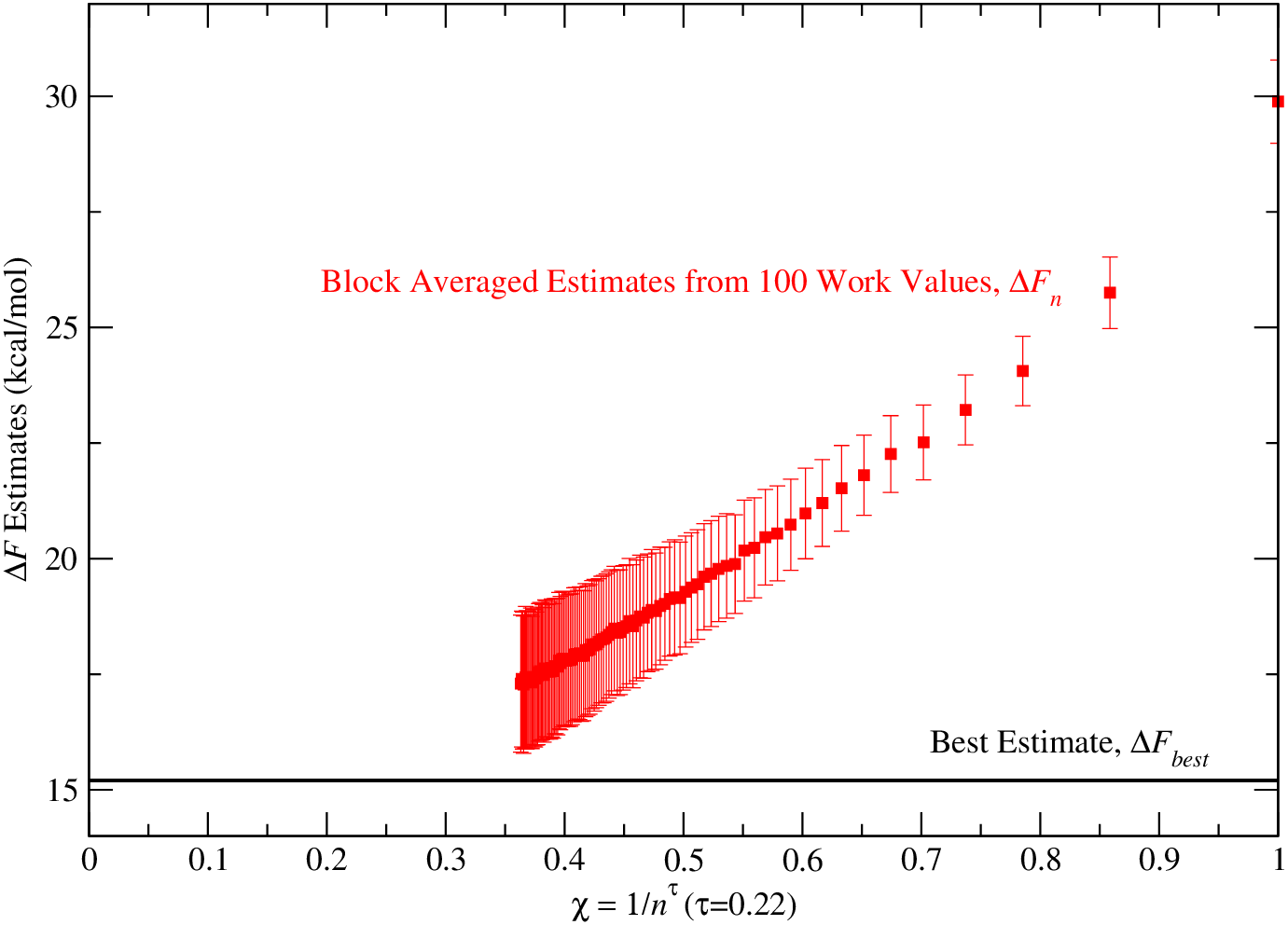}
  \caption{
    \label{fig-linear}
    Bootstrapped block averaged free energy ($\Delta F_n$),
    given by Eq.\ (\ref{eq-block})
    as a function of $\chi$ are shown as red
    squares. The solid black line represents the
    best estimate $\Delta F_{best}$. 
    The value of $\tau=0.22$ is chosen to minimize
    the slope of $\Delta F_n(\chi)$ as described in
    Sec.\ \ref{sec-linear}.
    This plot was generated using 100 work values
    chosen at random from the PAL2STE test system (palmitic $\rightarrow$
    stearic acid mutation).
    In this plot, extrapolating to the infinite data limit
    cooresponds to continuing the $\Delta F_n$ curve to $\chi=0$
    to obtain the intercept.
    This plot also demonstrates that the large $\chi$ (small $n$)
    data are more reliable as shown by the errorbars which
    represent the standard error of $\Delta F_n$.
  }
\end{figure}

Now that a smooth function has been obtained in the block averaged
free energy $\Delta F_n$,
shown in Fig.\ \ref{fig-block}, extrapolation to the
infinite data limit becomes feasible, as originally suggested
by Zuckerman and Woolf \cite{zuckerman-cpl}.
The basic idea is to plot $\Delta F_n$ as 
a function of some variable and
then extrapolate to the infinite
data limit ($n \rightarrow \infty$).
It is useful to plot $\Delta F_n$ as a function of
$\chi=1/n^\tau$ as shown
in Fig.\ \ref{fig-linear} \cite{zuckerman-cpl}.
The plot was generated
by choosing 100 work values at random from the PAL2STE data set.
$\Delta F_n$ was then computed for this subset of 100 work values
following the steps outlined in Sec.\ \ref{sec-block}.
In this plot the {\it bootstrapped} $\Delta F_n$ is shown,
and the best estimate $\Delta F_{best}$ is included
as the solid black line.
A value of $\tau=0.22$ was chosen to minimize the slope
of $\Delta F_n(\chi)$ as discussed below.
The errorbars show the statistical uncertainty of the
$\Delta F_n$ given by the standard error associated with $\sigma_n$
in Eq.\ (\ref{eq-blockSn}). The smallest uncertainty occurs
for $\chi=n=1$ due to the fact that $\Delta F_n(\chi=1)$ is simply
the average work.

It is useful to plot $\Delta F_n$ as a function of $\chi=1/n^\tau$
as in Fig.\ \ref{fig-linear} (rather than $n$) because
the infinite data limit ($n \rightarrow \infty$)
now corresponds to $\chi=0$.
In addition, this simple form gives a bounded interval ($\chi=(0,1]$),
rather than an infinite one (such as $\Delta F_n$ as a function of $n$).
This form allows us to develop two simple extrapolation
schemes as explained in the following sections.

\subsection{\label{sec-linear} Linear Extrapolation}
It is known that the block averaged
free energy $\Delta F_n$ in Eq.\ (\ref{eq-block})
guarantees monotonic behavior
\cite{zuckerman-cpl,zuckerman-prl,zuckerman-jstat}.
Thus, one can hope to obtain
a reasonable estimate of $\Delta F$
by simply continuing the curve
in Fig.\ \ref{fig-linear} with a straight line.
Such a linear extrapolation
guarantees that our extrapolated results will not exceed
$\Delta F_n$ for $n=N$ -- neccessary since
$\Delta F_n$ is a rigorous upper 
bound for the true $\Delta F$
\cite{zuckerman-jstat}.

We test this extrapolation method using the test systems
described in \ref{sec-systems}. This {\it fully automated}
process contains the following steps:
(i) Draw a subset containing $N$ work values
($W_1,W_2,...,W_N$) at
random from the data set.
(ii) Plot the {\it bootstrapped} $\Delta F_n$ as
a function of $\chi=1/n^{\tau}$; vary $\tau$, then
choose the value of $\tau$ that minimizes the slope of the
tail (i.e.\ small $\chi$) of $\Delta F_n$. (If one
has enough data to get the correct $\Delta F$ then, for
the right value of $\tau$, the slope will be nearly zero.)
(iii) Extrapolate $\Delta F_n$ to $\chi=0$ using a straight line. The
intercept ($\chi=0$) is our extrapolated free energy $\Delta F_{lin}$.
(iv) Using these same $N$ work values, estimate the free energy
$\Delta F_{Jarz}$ with Eq.\ (\ref{eq-jarz}).
This process is repeated 500 times to obtain the average and
standard deviation of our $\Delta F_{lin}$ and $\Delta F_{Jarz}$.

A simple extension of the linear method shown here, is to fit $\Delta F_n$
to a nonlinear function, such as quadratic in $\chi$, as
in previous work by Zuckerman and Woolf \cite{zuckerman-cpl}. These
nonlinear extrapolation methods offer little, if any, improvement
in the average
$\Delta F$ extrapolations. And, due to the inherent instability
of high order fits, the standard deviations
for the extrapolated results are much larger than those obtained
for linear extrapolation.

\subsection{\label{sec-rci} Reverse Cumulative Integral Extrapolation}
As previously metioned (see Fig.\ \ref{fig-linear}),
the most precise $\Delta F_n$ values
occur for larger
$\chi\approx 1$ (i.e.\ smaller $n$), yet the previous linear
extrapolation scheme relies exclusively on small $\chi$ values. Thus,
in an effort to use the
more precise large-$\chi$ data to extrapolate $\Delta F$, we now formulate
an integration scheme which explicity includes all values of $\chi$.

\begin{figure}
  \includegraphics[scale=0.5]{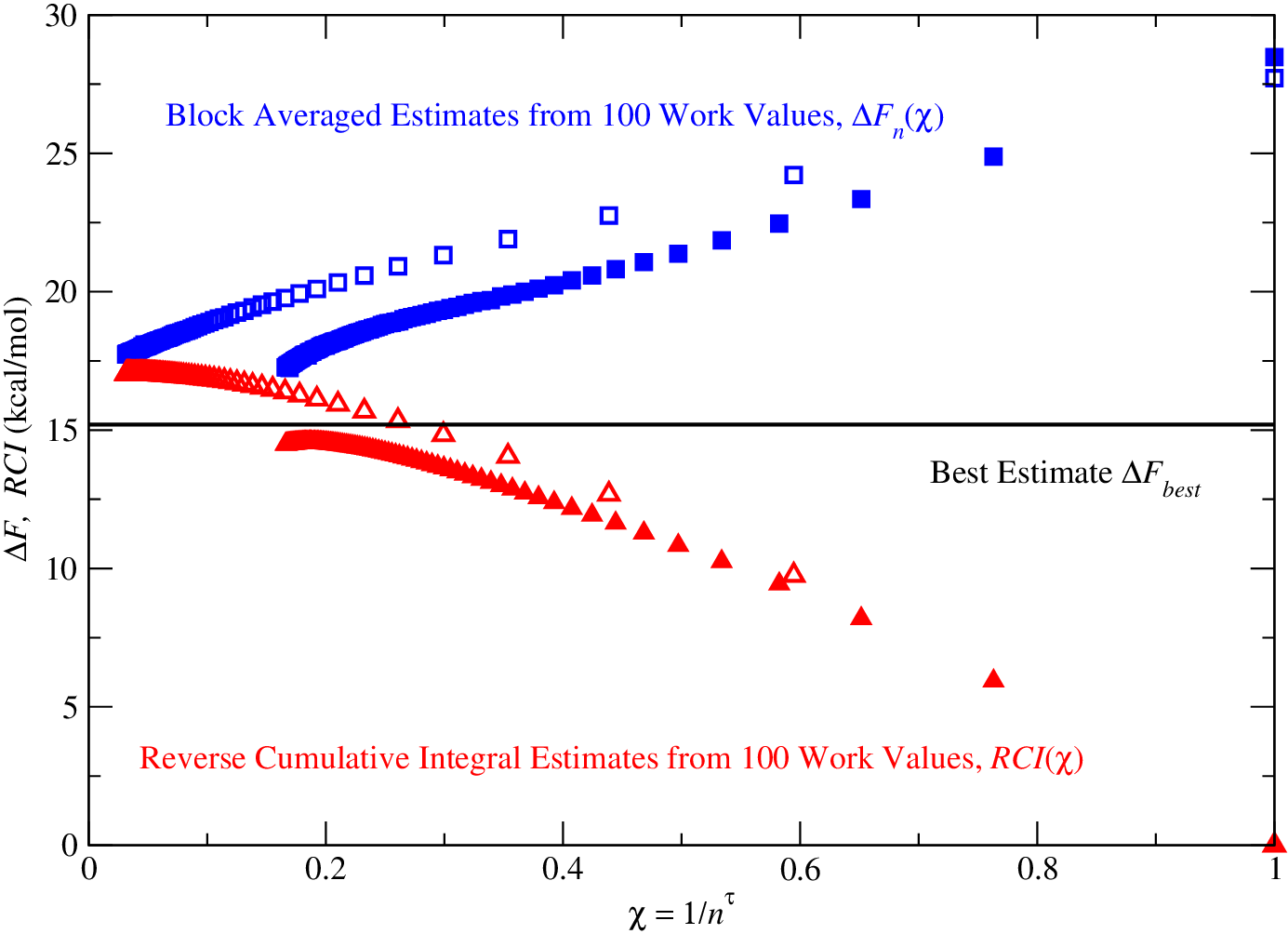}
  \caption{
    \label{fig-rci}
    Examples of the reverse cumulative integral, $RCI(\chi)$
    are shown for a two {\it subsets} of 100 work values drawn
    at random from the PAL2STE data set (palmitic to stearic acid
    mutation). The first subset is represented by
    open symbols and the second by closed symbols.
    The solid black line shows the
    best estimate $\Delta F_{best}$, the blue squares are
    the sub-sampled
    $\Delta F_n$ and the red circles are the $RCI(\chi$).
    The strength of using $RCI(\chi)$ for extrapolation
    is its explicit use of all the $\Delta F_n$ values.
    For each subset, the value of $\tau$ was chosen to
    minimize the
    slope of the small-$\chi$ tail of $RCI(\chi)$, as
    described in Sec.\ \ref{sec-rci}.
    In this example, the subset represented by the open
    symbols slightly overestimates $\Delta F_{best}$,
    while the subset represented by the closed symbols
    slightly underestimates $\Delta F_{best}$.
  }
\end{figure}

Consider treating $\Delta F_n$ in Fig.\ \ref{fig-linear}
as a smooth function $\Delta F_n(\chi)$, from $\chi=0$ to 1. We are
free to consider the area under this function, re-written using
integration by parts,
\begin{eqnarray}
  \int_0^1 d\chi \Delta F_n(\chi)=
  \int_0^1 d\chi(1-\chi)\frac{d\Delta F_n(\chi)}{d\chi}
  +\Delta F_n(\chi=0).
\end{eqnarray}
But $\Delta F_n(\chi=0)$ is just the extrapolated
free energy estimate $\Delta F_{rci}$, so
\begin{eqnarray}
  \Delta F_{rci}=\int_0^1 d\chi\Biggl(\Delta F_n(\chi)-
  (1-\chi)\frac{d\Delta F_n(\chi)}{d\chi}\Biggr).
\end{eqnarray}
Now the {\it reverse cumulative integral} function can be defined by
\begin{eqnarray}
  RCI(\chi)=\int_1^\chi d\chi'\Biggl(\Delta F_n(\chi')-
  (1-\chi')\frac{d\Delta F_n(\chi')}{d\chi'}\Biggr),
\end{eqnarray}
where it should be noted that we accumulate in the
reverse direction from $\chi'=1$, where
the data is most precise, to $\chi'=\chi$, i.e.\ from right to left
in Figs.\ \ref{fig-linear} and \ref{fig-rci}.

A sample plot of the reverse cumulative integral is
shown in Fig.\ \ref{fig-rci}. This plot was generated
using two {\it subsets} (represented by
open and closed symbols)
of 100 work values drawn at random from the
PAL2STE data set. The solid black line shows the
best estimate $\Delta F_{best}$,
the blue squares are the sub-sampled
$\Delta F_n$ and the red circles are $RCI(\chi$). For
each of the two subsets, the value of $\tau$ was
chosen to minimize the slope of the tail of $RCI(\chi)$,
as discussed below. The subset represented by
the open symbols slightly overestimates $\Delta F_{best}$,
while the subset represented by the closed symbols
slightly underestimates $\Delta F_{best}$.

To obtain an extrapolated value for $\Delta F$, consider the
case where one has more than enough
data to obtain $\Delta F$ exactly.
In this situation, {\it if $\tau$ is chosen carefully},
$RCI(\chi)$ will have nearly zero slope for small $\chi$, since
accumulating more $\chi$ values will not change the estimate.
Thus, one can hope to extrapolate $\Delta F$ by simply finding
a value of $\tau$ where the slope $dRCI(\chi)/d\chi$ is
the smallest for small $\chi$, then
the extrapolated free energy $\Delta F_{rci}$ will be
the value of $RCI(\chi)$ for the smallest value of $\chi$ available,
$\chi_{min}$.

Our fully automated test of this new extrapolation
method is very similar to that 
described in the previous section, with only minor differences:
(i) the {\it sub-sampled} $\Delta F_n$ is used,
(ii) the value of $\tau$ is chosen to minimize the slope
of the tail (small $\chi$) of $RCI(\chi)$ -- see Fig.\
\ref{fig-rci}, and
(iii) once the value of $\tau$ is determined, the free
energy is estimated to by $\Delta F_{rci}=RCI(\chi_{min})$.
Comparison is made with the Jarzynksi estimate
$\Delta F_{Jarz}$ using the same procedure
as in the last section.

\section{\label{sec-results} Results}

\begin{figure}
  \includegraphics[scale=0.65]{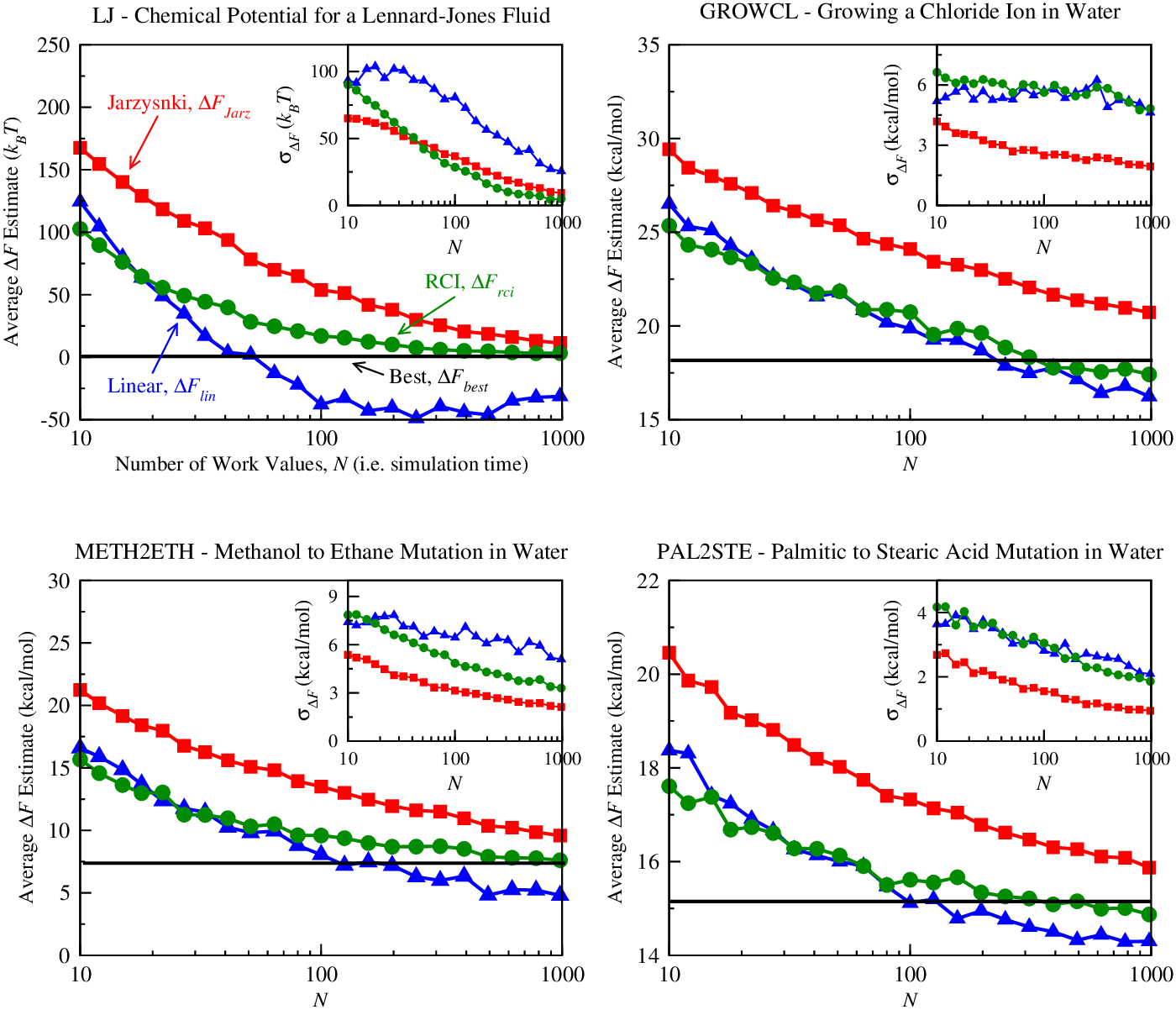}
  \caption{
    \label{fig-extrap}
    A comparison between $\Delta F$ estimates for
    linear extrapolation, reverse
    cumulative integral (RCI) extrapolation, and the
    Jarzynski equality for all of the test systems.
    For each of the plots the
    solid horizontal black line indicates the best estimate
    $\Delta F_{best}$ given in Sec.\ \ref{sec-systems},
    the red squares are averages of Jarzynski
    estimates given by Eq.\ (\ref{eq-jarz}), the
    blue triangles are averages of linearly extrapolated
    estimates from Sec.\ \ref{sec-linear}, and the green
    circles are averages of RCI extrapolated estimates from
    Sec.\ \ref{sec-rci}.
    The inset in each plot shows the standard deviation
    for each of the estimates.
    Averages and stardard deviations were
    obtained by performing 500 independent trials for
    each estimate for each value of $N$.
  }
\end{figure}

The initial results of this study are
very positive as shown by the rapid convergence
of our extrapolated $\Delta F$ estimates
(Fig.\  \ref{fig-extrap}).
Compared to $\Delta F_{Jarz}$, estimates of $\Delta F$
can be made with 6-15 fold less work values,
i.e.\ less computational expense.

Fig.\ \ref{fig-extrap} demonstrates how the linear
and reverse cumulative extrapolation (RCI) methods
described above compare
to using the Jarzynski estimate of Eq.\ (\ref{eq-jarz}),
for each of the four test systems.
For all of the plots shown, the solid black horizontal 
line corresponds to the
Jarzynski estimate using all available work values and thus
represents the best estimate $\Delta F_{best}$ from
Sec.\ \ref{sec-systems}. The red squares are 
averages of $\Delta F_{Jarz}$ using Eq.\ (\ref{eq-jarz}),
the blue triangles are
averages of $\Delta F_{lin}$ from Sec.\ \ref{sec-linear},
and the green circles are averages of
$\Delta F_{rci}$ from Sec.\ \ref{sec-rci}. The inset
for each plot shows the standard deviation
of the $\Delta F$ estimates ($\sigma_{\Delta F}$).
Averages and stardard deviations were
obtained by performing 500 independent trials for
each estimate
($\Delta F_{Jarz}$, $\Delta F_{lin}$, $\Delta F_{rci}$)
for every value of $N$. Thus $\sigma_{\Delta F}$ indicates
the expected statistical
uncertainty -- that is the range of values one would expect if the
calculation was performed {\it de novo}.

A glance at Fig.\ \ref{fig-extrap} reveals
that the linearly extrapolated $\Delta F_{lin}$ estimates
converge to the best estimate $\Delta F_{best}$
more quickly than the Jarzynski estimate $\Delta F_{Jarz}$.
The larger uncertainty of the linearly extrapolated
estimates is, at least partially, explained by
the fact that it relies on the less certain $\Delta F_n$
values as explained in Sec.\ \ref{sec-linear}. Also, the
linear estimates tend to ``overshoot'' $\Delta F_{best}$.

Many of the disadvantages of the linearly extrapolated
estimates are somewhat overcome by RCI extrapolation.
Since RCI extrapolation relies heavily on the more
precise values of
$\Delta F_n$, the uncertainty is generally smaller than that
of the linear estimates. Remarkably, for the LJ system,
the RCI extrapolated uncertainty is smaller than the Jarzysnki
estimate uncertainty for $N>40$. Also, RCI extrapolated estimates
do not tend to appreciably overshoot $\Delta F_{best}$.

\begin{table}
  \caption{\label{tab-results} A quantitative comparison
    between the reverse cumulative integral
    estimates ($\Delta F_{rci}$)
    and the Jarzysnki estimate ($\Delta F_{Jarz}$)
    shown in Fig.\ \ref{fig-extrap}.
    The first column shows the test system used in the
    comparison.
    The second and third columns are the number of work
    values needed to obtain an estimate that falls within
    1.0 kcal/mol of $\Delta F_{best}$ for the reverse
    cumulative integral
    ($N_{rci}$) and Jarzynksi ($N_{Jarz}$) estimates. The
    rightmost column is the ratio of these two values, i.e.\
    the approximate improvement of the reverse cumulative
    integral estimate over the Jarzysnki estimate.}
    \begin{tabular}{lccc}
      \hline
      System   &\,\, $N_{rci}$\,\, &\,\, $N_{Jarz}$ \,\,& Improvement\\
      \hline
      LJ       & 800               & 6000               & 7.5        \\
      GROWCL   & 200               & 3000               & 15         \\
      METH2ETH & 400               & 2500               & 6.25       \\
      PAL2STE  & 40                & 500                & 12.5       \\
      \hline
    \end{tabular}
\end{table}

To obtain a quantitative comparison between
RCI extrapolated estimates $\Delta F_{rci}$,
and Jarzynski estimates $\Delta F_{Jarz}$,
we ask the following question: how many work values are necessary
to obtain a $\Delta F$ estimate that falls within 1.0 kcal/mol 
of the best estimate $\Delta F_{best}$?
Table \ref{tab-results} summarizes the results of this
comparison. The RCI estimates offer a significant improvement
over the Jarzynski estimates in all of the test systems, with
a 6-15 fold decrease in the number of work values needed
to estimate $\Delta F_{best}$ within 1.0 kcal/mol.

Due to the fact that the linearly extrapolated estimates tend
to overshoot $\Delta F_{best}$, often by many kcal/mol,
a quantitative comparison
between $\Delta F_{lin}$ and $\Delta F_{Jarz}$ would be 
difficult and unreliable. Thus, we do not attempt to make
such a comparison here.

\section{Conclusion}

We have described two methods that
improve standard non-equilibrium estimates of
free energy differences,
$\Delta F$: linear extrapolation and
reverse cumulative integral (RCI) extrapolation. 
Four test systems were used in this study: chemical
potential calculation for a Lennard-Jones fluid,
growing a chloride ion in water, methanol $\rightarrow$
ethane mutation in water, and palmitic $\rightarrow$ stearic
acid mutation in water. Both of the methods
rely on block averaged free energies $\Delta F_n$,
which are extrapolated to the infinite data limit, 
and offer more rapid estimates of $\Delta F$ than using
the Jarzynski equality alone,
for the test systems considered here.

Previous work by Zuckerman and Woolf
\cite{zuckerman-cpl} used a quadratic
extrapolation method to estimate $\Delta F$. The
present study offers several improvements:
(i) the accuracy and uncertainty of the extrapolated
estimates are reduced due to improved, fully automated
methods;
(ii) two new methods for calculating the block
averaged free energies, $\Delta F_n$ are described;
(iii) a key innovation is offered in RCI extrapolation
in its use of the more reliable $\Delta F_n$ data;
(iv) a systematic quantitative comparison is done between
the RCI and Jarzynksi $\Delta F$ estimates, showing a
6-15 fold decrease in the number of work values
needed for the RCI estimates;
(v) we have tested our extrapolation methods on four systems of
varying molecular complexity. 

For the first time, bootstrapped and
sub-sampled block averaged free energies are
introduced. These $\Delta F_n$ offer very smooth
convergence properties allowing statistically
reliable extrapolation. The ability to generate smooth
$\Delta F_n$ data is critical to the extrapolation
methods described here.

A quantitative comparison between the
RCI extrapolated $\Delta F$ estimates and those
using Jarzynski's equality show a marked
decrease in the number of work
values needed to estimate $\Delta F$ when
using the RCI estimates. RCI extrapolation
can obtain $\Delta F$ estimates
using 6-15 fold less data than the Jarzysnki estimates.
The linear extrapolation estimates tend to overshoot the best
estimate $\Delta F$ and has a larger uncertainty than
RCI extrapolation. However, the partial success of the
``simple-minded'' linear extrapolation does illustrate
the power of the underlying idea:
systematic behavior in bias can be exploited.

Other similar extrapolation methods could be developed
that may offer improvement over those presented here.
Such methods are currenty under investigation by the authors.
Future work by the authors will use extrapolation methods,
such as those described here, to generate $\Delta F$ estimates
for large molecular systems such as relative
protein-ligand binding affinities.

\section{Acknowledgements}
We gratefully acknowledge Jay Ponder for
granting F.\ M.\ Y.\ permission to write
code for TINKER to perform fast-growth free energy
calculations, and
to Alan Grossfield for his assistance in writing
and implementing this code. We would like to
thank Arun Setty for fruitful discussion
and valuable suggestions. Special thanks are due to
Hirsh Nanda and Thomas Woolf for permission to use
the raw data for the palmitic to stearic mutation,
and to David Hendrix and Chris Jarzysnki for permission
to use the raw data for the Lennard-Jones fluid.
Funding for this research was provided by
the Dept.\ of Environmental and Occupational
Health at University of Pittsburgh, and
the National Institutes of Health (T32ES007318).

\bibliography{my}

\begin{thebibliography}{65}
\expandafter\ifx\csname natexlab\endcsname\relax\def\natexlab#1{#1}\fi
\expandafter\ifx\csname bibnamefont\endcsname\relax
  \def\bibnamefont#1{#1}\fi
\expandafter\ifx\csname bibfnamefont\endcsname\relax
  \def\bibfnamefont#1{#1}\fi
\expandafter\ifx\csname citenamefont\endcsname\relax
  \def\citenamefont#1{#1}\fi
\expandafter\ifx\csname url\endcsname\relax
  \def\url#1{\texttt{#1}}\fi
\expandafter\ifx\csname urlprefix\endcsname\relax\def\urlprefix{URL }\fi
\providecommand{\bibinfo}[2]{#2}
\providecommand{\eprint}[2][]{\url{#2}}

\bibitem[{\citenamefont{Zuckerman and
  Woolf}(2002{\natexlab{a}})}]{zuckerman-cpl}
\bibinfo{author}{\bibfnamefont{D.~M.} \bibnamefont{Zuckerman}}
  \bibnamefont{and} \bibinfo{author}{\bibfnamefont{T.~B.} \bibnamefont{Woolf}},
  \bibinfo{journal}{Chem.\ Phys.\ Lett.} \textbf{\bibinfo{volume}{351}},
  \bibinfo{pages}{445} (\bibinfo{year}{2002}{\natexlab{a}}).

\bibitem[{\citenamefont{Zuckerman and
  Woolf}(2002{\natexlab{b}})}]{zuckerman-prl}
\bibinfo{author}{\bibfnamefont{D.~M.} \bibnamefont{Zuckerman}}
  \bibnamefont{and} \bibinfo{author}{\bibfnamefont{T.~B.} \bibnamefont{Woolf}},
  \bibinfo{journal}{Phys.\ Rev.\ Lett.} \textbf{\bibinfo{volume}{89}},
  \bibinfo{pages}{180602} (\bibinfo{year}{2002}{\natexlab{b}}).

\bibitem[{\citenamefont{Shirts et~al.}(2003{\natexlab{a}})\citenamefont{Shirts,
  Bair, Hooker, and Pande}}]{shirts-prl}
\bibinfo{author}{\bibfnamefont{M.~R.} \bibnamefont{Shirts}},
  \bibinfo{author}{\bibfnamefont{E.}~\bibnamefont{Bair}},
  \bibinfo{author}{\bibfnamefont{G.}~\bibnamefont{Hooker}}, \bibnamefont{and}
  \bibinfo{author}{\bibfnamefont{V.~S.} \bibnamefont{Pande}},
  \bibinfo{journal}{Phys.\ Rev.\ Lett.} \textbf{\bibinfo{volume}{91}},
  \bibinfo{pages}{140601} (\bibinfo{year}{2003}{\natexlab{a}}).

\bibitem[{\citenamefont{Pearlman and Kollman}(1989{\natexlab{a}})}]{pearlman}
\bibinfo{author}{\bibfnamefont{D.~A.} \bibnamefont{Pearlman}} \bibnamefont{and}
  \bibinfo{author}{\bibfnamefont{P.~A.} \bibnamefont{Kollman}},
  \bibinfo{journal}{J.\ Chem.\ Phys.} \textbf{\bibinfo{volume}{90}},
  \bibinfo{pages}{2460} (\bibinfo{year}{1989}{\natexlab{a}}).

\bibitem[{\citenamefont{Kong and Brooks}(1996)}]{kong}
\bibinfo{author}{\bibfnamefont{X.}~\bibnamefont{Kong}} \bibnamefont{and}
  \bibinfo{author}{\bibfnamefont{C.~L.} \bibnamefont{Brooks}},
  \bibinfo{journal}{J.\ Chem.\ Phys.} \textbf{\bibinfo{volume}{105}},
  \bibinfo{pages}{2414} (\bibinfo{year}{1996}).

\bibitem[{\citenamefont{Shirts et~al.}(2003{\natexlab{b}})\citenamefont{Shirts,
  Pitera, Swope, and Pande}}]{shirts-jcp}
\bibinfo{author}{\bibfnamefont{M.~R.} \bibnamefont{Shirts}},
  \bibinfo{author}{\bibfnamefont{J.~W.} \bibnamefont{Pitera}},
  \bibinfo{author}{\bibfnamefont{W.~C.} \bibnamefont{Swope}}, \bibnamefont{and}
  \bibinfo{author}{\bibfnamefont{V.~S.} \bibnamefont{Pande}},
  \bibinfo{journal}{J.\ Chem.\ Phys.} \textbf{\bibinfo{volume}{119}},
  \bibinfo{pages}{5740} (\bibinfo{year}{2003}{\natexlab{b}}).

\bibitem[{\citenamefont{Hummer and Szabo}(2001)}]{hummer-pull}
\bibinfo{author}{\bibfnamefont{G.}~\bibnamefont{Hummer}} \bibnamefont{and}
  \bibinfo{author}{\bibfnamefont{A.}~\bibnamefont{Szabo}},
  \bibinfo{journal}{Proc.\ Nat.\ Acad.\ Sci.\ (USA)}
  \textbf{\bibinfo{volume}{98}}, \bibinfo{pages}{3658} (\bibinfo{year}{2001}).

\bibitem[{\citenamefont{Bash et~al.}(1987)\citenamefont{Bash, Singh, Langridge,
  and Kollman}}]{bash-sci}
\bibinfo{author}{\bibfnamefont{P.~A.} \bibnamefont{Bash}},
  \bibinfo{author}{\bibfnamefont{U.~C.} \bibnamefont{Singh}},
  \bibinfo{author}{\bibfnamefont{R.}~\bibnamefont{Langridge}},
  \bibnamefont{and} \bibinfo{author}{\bibfnamefont{P.~A.}
  \bibnamefont{Kollman}}, \bibinfo{journal}{Science}
  \textbf{\bibinfo{volume}{236}}, \bibinfo{pages}{564} (\bibinfo{year}{1987}).

\bibitem[{\citenamefont{Woods et~al.}(2003)\citenamefont{Woods, Essex, and
  King}}]{woods-jpcB}
\bibinfo{author}{\bibfnamefont{C.~J.} \bibnamefont{Woods}},
  \bibinfo{author}{\bibfnamefont{J.~W.} \bibnamefont{Essex}}, \bibnamefont{and}
  \bibinfo{author}{\bibfnamefont{M.~A.} \bibnamefont{King}},
  \bibinfo{journal}{J.\ Phys.\ Chem.\ B} \textbf{\bibinfo{volume}{107}},
  \bibinfo{pages}{13703} (\bibinfo{year}{2003}).

\bibitem[{\citenamefont{Mc{C}ammon}(1991)}]{mccammon}
\bibinfo{author}{\bibfnamefont{J.~A.} \bibnamefont{Mc{C}ammon}},
  \bibinfo{journal}{Curr.\ Opin.\ Struc.\ Bio.} \textbf{\bibinfo{volume}{2}},
  \bibinfo{pages}{96} (\bibinfo{year}{1991}).

\bibitem[{\citenamefont{Shobana et~al.}(2000)\citenamefont{Shobana, Roux, and
  Andersen}}]{shobana}
\bibinfo{author}{\bibfnamefont{S.}~\bibnamefont{Shobana}},
  \bibinfo{author}{\bibfnamefont{B.}~\bibnamefont{Roux}}, \bibnamefont{and}
  \bibinfo{author}{\bibfnamefont{O.~S.} \bibnamefont{Andersen}},
  \bibinfo{journal}{J.\ Phys.\ Chem.\ B} \textbf{\bibinfo{volume}{104}},
  \bibinfo{pages}{5179} (\bibinfo{year}{2000}).

\bibitem[{\citenamefont{Isralewitz et~al.}(2001)\citenamefont{Isralewitz, Gao,
  and Schulten}}]{isralewitz}
\bibinfo{author}{\bibfnamefont{B.}~\bibnamefont{Isralewitz}},
  \bibinfo{author}{\bibfnamefont{M.}~\bibnamefont{Gao}}, \bibnamefont{and}
  \bibinfo{author}{\bibfnamefont{K.}~\bibnamefont{Schulten}},
  \bibinfo{journal}{Curr.\ Opin.\ Struc.\ Bio.} \textbf{\bibinfo{volume}{11}},
  \bibinfo{pages}{224} (\bibinfo{year}{2001}).

\bibitem[{\citenamefont{Bitetti-Putzer
  et~al.}(2003)\citenamefont{Bitetti-Putzer, Yang, and Karplus}}]{karplus-cpl}
\bibinfo{author}{\bibfnamefont{R.}~\bibnamefont{Bitetti-Putzer}},
  \bibinfo{author}{\bibfnamefont{W.}~\bibnamefont{Yang}}, \bibnamefont{and}
  \bibinfo{author}{\bibfnamefont{M.}~\bibnamefont{Karplus}},
  \bibinfo{journal}{Chem.\ Phys.\ Lett.} \textbf{\bibinfo{volume}{377}},
  \bibinfo{pages}{633} (\bibinfo{year}{2003}).

\bibitem[{\citenamefont{Boresch et~al.}(2003)\citenamefont{Boresch, Tettinger,
  Leitgeb, and Karplus}}]{karplus-jpcB}
\bibinfo{author}{\bibfnamefont{S.}~\bibnamefont{Boresch}},
  \bibinfo{author}{\bibfnamefont{F.}~\bibnamefont{Tettinger}},
  \bibinfo{author}{\bibfnamefont{M.}~\bibnamefont{Leitgeb}}, \bibnamefont{and}
  \bibinfo{author}{\bibfnamefont{M.}~\bibnamefont{Karplus}},
  \bibinfo{journal}{J.\ Phys.\ Chem.\ B} \textbf{\bibinfo{volume}{107}},
  \bibinfo{pages}{9535} (\bibinfo{year}{2003}).

\bibitem[{\citenamefont{Liphardt et~al.}(2002)\citenamefont{Liphardt, Dumont,
  Smith, Tinoco, and Bustamante}}]{bustamante-sci}
\bibinfo{author}{\bibfnamefont{J.}~\bibnamefont{Liphardt}},
  \bibinfo{author}{\bibfnamefont{S.}~\bibnamefont{Dumont}},
  \bibinfo{author}{\bibfnamefont{S.~B.} \bibnamefont{Smith}},
  \bibinfo{author}{\bibfnamefont{I.}~\bibnamefont{Tinoco}}, \bibnamefont{and}
  \bibinfo{author}{\bibfnamefont{C.}~\bibnamefont{Bustamante}},
  \bibinfo{journal}{Science} \textbf{\bibinfo{volume}{296}},
  \bibinfo{pages}{1832} (\bibinfo{year}{2002}).

\bibitem[{\citenamefont{Gore et~al.}(2003)\citenamefont{Gore, Ritort, and
  Bustamante}}]{bustamante-pnas}
\bibinfo{author}{\bibfnamefont{J.}~\bibnamefont{Gore}},
  \bibinfo{author}{\bibfnamefont{J.}~\bibnamefont{Ritort}}, \bibnamefont{and}
  \bibinfo{author}{\bibfnamefont{C.}~\bibnamefont{Bustamante}},
  \bibinfo{journal}{Proc.\ Natl.\ Acad.\ Sci.\ (USA)}
  \textbf{\bibinfo{volume}{100}}, \bibinfo{pages}{12564}
  (\bibinfo{year}{2003}).

\bibitem[{\citenamefont{Park et~al.}(2003)\citenamefont{Park, Khalili-Araghi,
  Tajkhorshid, and Schulten}}]{schulten-2003}
\bibinfo{author}{\bibfnamefont{S.}~\bibnamefont{Park}},
  \bibinfo{author}{\bibfnamefont{F.}~\bibnamefont{Khalili-Araghi}},
  \bibinfo{author}{\bibfnamefont{E.}~\bibnamefont{Tajkhorshid}},
  \bibnamefont{and} \bibinfo{author}{\bibfnamefont{K.}~\bibnamefont{Schulten}},
  \bibinfo{journal}{J.\ Chem.\ Phys.} \textbf{\bibinfo{volume}{119}},
  \bibinfo{pages}{3559} (\bibinfo{year}{2003}).

\bibitem[{\citenamefont{Bajorath}(2002)}]{bajorath}
\bibinfo{author}{\bibfnamefont{J.}~\bibnamefont{Bajorath}},
  \bibinfo{journal}{Nature Rev.\ Drug Disc.} \textbf{\bibinfo{volume}{1}},
  \bibinfo{pages}{882} (\bibinfo{year}{2002}).

\bibitem[{\citenamefont{Abraham}(2003)}]{burgers}
\bibinfo{editor}{\bibfnamefont{D.~J.} \bibnamefont{Abraham}}, ed.,
  \emph{\bibinfo{title}{Burger's Medicinal Chemistry and Drug Discovery, Sixth
  Ed., Volume 1}} (\bibinfo{publisher}{Wiley}, \bibinfo{address}{New York},
  \bibinfo{year}{2003}).

\bibitem[{\citenamefont{Lazar et~al.}(2003)\citenamefont{Lazar, Marshall,
  Plecs, Mayo, and Desjarlais}}]{mayo}
\bibinfo{author}{\bibfnamefont{G.~A.} \bibnamefont{Lazar}},
  \bibinfo{author}{\bibfnamefont{S.~A.} \bibnamefont{Marshall}},
  \bibinfo{author}{\bibfnamefont{J.~J.} \bibnamefont{Plecs}},
  \bibinfo{author}{\bibfnamefont{S.~L.} \bibnamefont{Mayo}}, \bibnamefont{and}
  \bibinfo{author}{\bibfnamefont{J.~R.} \bibnamefont{Desjarlais}},
  \bibinfo{journal}{Curr.\ Op.\ Struct.\ Bio.} \textbf{\bibinfo{volume}{13}},
  \bibinfo{pages}{513} (\bibinfo{year}{2003}).

\bibitem[{\citenamefont{De{G}rado and Nilsson}(1997)}]{degrado}
\bibinfo{author}{\bibfnamefont{W.~F.} \bibnamefont{De{G}rado}}
  \bibnamefont{and} \bibinfo{author}{\bibfnamefont{B.~O.}
  \bibnamefont{Nilsson}}, \bibinfo{journal}{Curr.\ Op.\ Struct.\ Bio.}
  \textbf{\bibinfo{volume}{7}}, \bibinfo{pages}{455} (\bibinfo{year}{1997}).

\bibitem[{\citenamefont{Jarzynski}(1997{\natexlab{a}})}]{jarzynski-both}
\bibinfo{author}{\bibfnamefont{C.}~\bibnamefont{Jarzynski}},
  \bibinfo{journal}{Phys.\ Rev.\ Lett.} \textbf{\bibinfo{volume}{78}},
  \bibinfo{pages}{2690} (\bibinfo{year}{1997}{\natexlab{a}}),
  \bibinfo{note}{{P}hys.\ Rev.\ E {\bf 56}, 5018 (1997)}.

\bibitem[{\citenamefont{Schurr and Fujimoto}(2002)}]{schurr}
\bibinfo{author}{\bibfnamefont{J.~M.} \bibnamefont{Schurr}} \bibnamefont{and}
  \bibinfo{author}{\bibfnamefont{B.~S.} \bibnamefont{Fujimoto}},
  \bibinfo{journal}{Science} \textbf{\bibinfo{volume}{296}},
  \bibinfo{pages}{1832} (\bibinfo{year}{2002}).

\bibitem[{\citenamefont{Zuckerman and Woolf}()}]{zuckerman-jstat}
\bibinfo{author}{\bibfnamefont{D.~M.} \bibnamefont{Zuckerman}}
  \bibnamefont{and} \bibinfo{author}{\bibfnamefont{T.~B.} \bibnamefont{Woolf}},
  \bibinfo{note}{{J}.\ Stat.\ Phys., in press.}

\bibitem[{\citenamefont{Grossfield et~al.}(2003)\citenamefont{Grossfield, Ren,
  and Ponder}}]{grossfield-jacs}
\bibinfo{author}{\bibfnamefont{A.}~\bibnamefont{Grossfield}},
  \bibinfo{author}{\bibfnamefont{P.}~\bibnamefont{Ren}}, \bibnamefont{and}
  \bibinfo{author}{\bibfnamefont{J.~W.} \bibnamefont{Ponder}},
  \bibinfo{journal}{J.\ Amer.\ Chem.\ Soc.} \textbf{\bibinfo{volume}{125}},
  \bibinfo{pages}{15671} (\bibinfo{year}{2003}).

\bibitem[{\citenamefont{Lu and Kofke}(2001{\natexlab{a}})}]{lu-i}
\bibinfo{author}{\bibfnamefont{N.}~\bibnamefont{Lu}} \bibnamefont{and}
  \bibinfo{author}{\bibfnamefont{D.~A.} \bibnamefont{Kofke}},
  \bibinfo{journal}{J.\ Chem.\ Phys.} \textbf{\bibinfo{volume}{114}},
  \bibinfo{pages}{7303} (\bibinfo{year}{2001}{\natexlab{a}}).

\bibitem[{\citenamefont{Lu and Kofke}(2001{\natexlab{b}})}]{lu-ii}
\bibinfo{author}{\bibfnamefont{N.}~\bibnamefont{Lu}} \bibnamefont{and}
  \bibinfo{author}{\bibfnamefont{D.~A.} \bibnamefont{Kofke}},
  \bibinfo{journal}{J.\ Chem.\ Phys.} \textbf{\bibinfo{volume}{115}},
  \bibinfo{pages}{6866} (\bibinfo{year}{2001}{\natexlab{b}}).

\bibitem[{\citenamefont{Oostenbrink and van Gunsteren}(2003)}]{oostenbrink}
\bibinfo{author}{\bibfnamefont{C.}~\bibnamefont{Oostenbrink}} \bibnamefont{and}
  \bibinfo{author}{\bibfnamefont{W.~F.} \bibnamefont{van Gunsteren}},
  \bibinfo{journal}{J.\ Comp.\ Chem.} \textbf{\bibinfo{volume}{24}},
  \bibinfo{pages}{1730} (\bibinfo{year}{2003}).

\bibitem[{\citenamefont{Mordasini and Mc{C}ammon}(2000)}]{mordasini}
\bibinfo{author}{\bibfnamefont{T.~Z.} \bibnamefont{Mordasini}}
  \bibnamefont{and} \bibinfo{author}{\bibfnamefont{J.~A.}
  \bibnamefont{Mc{C}ammon}}, \bibinfo{journal}{J.\ Phys.\ Chem.\ B}
  \textbf{\bibinfo{volume}{104}}, \bibinfo{pages}{360} (\bibinfo{year}{2000}).

\bibitem[{\citenamefont{Jarzynski}(2002)}]{jarzynski-targeted}
\bibinfo{author}{\bibfnamefont{C.}~\bibnamefont{Jarzynski}},
  \bibinfo{journal}{Phys.\ Rev.\ E} \textbf{\bibinfo{volume}{65}},
  \bibinfo{pages}{046122} (\bibinfo{year}{2002}).

\bibitem[{\citenamefont{Wood}(1991)}]{wood}
\bibinfo{author}{\bibfnamefont{R.~H.} \bibnamefont{Wood}},
  \bibinfo{journal}{J.\ Phys.\ Chem.} \textbf{\bibinfo{volume}{95}},
  \bibinfo{pages}{4838} (\bibinfo{year}{1991}).

\bibitem[{\citenamefont{Jarzynski}(1997{\natexlab{b}})}]{jarzynski-master}
\bibinfo{author}{\bibfnamefont{C.}~\bibnamefont{Jarzynski}},
  \bibinfo{journal}{Phys.\ Rev.\ E} \textbf{\bibinfo{volume}{56}},
  \bibinfo{pages}{5018} (\bibinfo{year}{1997}{\natexlab{b}}).

\bibitem[{\citenamefont{Miller and Reinhardt}(2000)}]{miller}
\bibinfo{author}{\bibfnamefont{M.~A.} \bibnamefont{Miller}} \bibnamefont{and}
  \bibinfo{author}{\bibfnamefont{W.~P.} \bibnamefont{Reinhardt}},
  \bibinfo{journal}{J.\ Chem.\ Phys.} \textbf{\bibinfo{volume}{113}},
  \bibinfo{pages}{7035} (\bibinfo{year}{2000}).

\bibitem[{\citenamefont{Hu et~al.}(2002)\citenamefont{Hu, Yun, and
  Hermans}}]{hu-ms}
\bibinfo{author}{\bibfnamefont{H.}~\bibnamefont{Hu}},
  \bibinfo{author}{\bibfnamefont{R.~H.} \bibnamefont{Yun}}, \bibnamefont{and}
  \bibinfo{author}{\bibfnamefont{J.}~\bibnamefont{Hermans}},
  \bibinfo{journal}{Molec.\ Sim.} \textbf{\bibinfo{volume}{28}},
  \bibinfo{pages}{67} (\bibinfo{year}{2002}).

\bibitem[{\citenamefont{Sch{\"{o}}n}(1996)}]{schon}
\bibinfo{author}{\bibfnamefont{J.~C.} \bibnamefont{Sch{\"{o}}n}},
  \bibinfo{journal}{J.\ Chem.\ Phys.} \textbf{\bibinfo{volume}{105}},
  \bibinfo{pages}{10072} (\bibinfo{year}{1996}).

\bibitem[{\citenamefont{Hodel et~al.}(1993)\citenamefont{Hodel, Simonson, Fox,
  and Br{\"u}nger}}]{brunger}
\bibinfo{author}{\bibfnamefont{A.}~\bibnamefont{Hodel}},
  \bibinfo{author}{\bibfnamefont{T.}~\bibnamefont{Simonson}},
  \bibinfo{author}{\bibfnamefont{R.~O.} \bibnamefont{Fox}}, \bibnamefont{and}
  \bibinfo{author}{\bibfnamefont{A.~T.} \bibnamefont{Br{\"u}nger}},
  \bibinfo{journal}{J.\ Phys.\ Chem.} \textbf{\bibinfo{volume}{97}},
  \bibinfo{pages}{3409} (\bibinfo{year}{1993}).

\bibitem[{\citenamefont{Pearlman and
  Kollman}(1989{\natexlab{b}})}]{pearlman-lag}
\bibinfo{author}{\bibfnamefont{D.~A.} \bibnamefont{Pearlman}} \bibnamefont{and}
  \bibinfo{author}{\bibfnamefont{P.~A.} \bibnamefont{Kollman}},
  \bibinfo{journal}{J.\ Chem.\ Phys.} \textbf{\bibinfo{volume}{91}},
  \bibinfo{pages}{7831} (\bibinfo{year}{1989}{\natexlab{b}}).

\bibitem[{\citenamefont{Di~{N}ola and Brunger}(1998)}]{dinola}
\bibinfo{author}{\bibfnamefont{A.}~\bibnamefont{Di~{N}ola}} \bibnamefont{and}
  \bibinfo{author}{\bibfnamefont{A.~T.} \bibnamefont{Brunger}},
  \bibinfo{journal}{J.\ Comp.\ Chem.} \textbf{\bibinfo{volume}{19}},
  \bibinfo{pages}{1229} (\bibinfo{year}{1998}).

\bibitem[{\citenamefont{Pearlman}(1994)}]{pearlman-1994}
\bibinfo{author}{\bibfnamefont{D.~A.} \bibnamefont{Pearlman}},
  \bibinfo{journal}{J.\ Comp.\ Chem.} \textbf{\bibinfo{volume}{15}},
  \bibinfo{pages}{105} (\bibinfo{year}{1994}).

\bibitem[{\citenamefont{Edholm and Ghosh}(1993)}]{edholm}
\bibinfo{author}{\bibfnamefont{O.}~\bibnamefont{Edholm}} \bibnamefont{and}
  \bibinfo{author}{\bibfnamefont{I.}~\bibnamefont{Ghosh}},
  \bibinfo{journal}{Molec.\ Sim.} \textbf{\bibinfo{volume}{10}},
  \bibinfo{pages}{241} (\bibinfo{year}{1993}).

\bibitem[{\citenamefont{Wood et~al.}(1991)\citenamefont{Wood, M{\"{u}}hlbauer,
  and Thompson}}]{wood-block}
\bibinfo{author}{\bibfnamefont{R.~H.} \bibnamefont{Wood}},
  \bibinfo{author}{\bibfnamefont{W.~C.~F.} \bibnamefont{M{\"{u}}hlbauer}},
  \bibnamefont{and} \bibinfo{author}{\bibfnamefont{P.~T.}
  \bibnamefont{Thompson}}, \bibinfo{journal}{J.\ Phys.\ Chem.}
  \textbf{\bibinfo{volume}{95}}, \bibinfo{pages}{6670} (\bibinfo{year}{1991}).

\bibitem[{\citenamefont{Lu et~al.}(2001)\citenamefont{Lu, Kofke, and
  Woolf}}]{lu-2001}
\bibinfo{author}{\bibfnamefont{N.}~\bibnamefont{Lu}},
  \bibinfo{author}{\bibfnamefont{D.~A.} \bibnamefont{Kofke}}, \bibnamefont{and}
  \bibinfo{author}{\bibfnamefont{T.~B.} \bibnamefont{Woolf}},
  \bibinfo{journal}{J.\ Chem.\ Phys.} \textbf{\bibinfo{volume}{115}},
  \bibinfo{pages}{6866} (\bibinfo{year}{2001}).

\bibitem[{\citenamefont{Lu et~al.}(2003)\citenamefont{Lu, Singh, and
  Kofke}}]{lu-2003}
\bibinfo{author}{\bibfnamefont{N.}~\bibnamefont{Lu}},
  \bibinfo{author}{\bibfnamefont{J.~K.} \bibnamefont{Singh}}, \bibnamefont{and}
  \bibinfo{author}{\bibfnamefont{D.~A.} \bibnamefont{Kofke}},
  \bibinfo{journal}{J.\ Chem.\ Phys.} \textbf{\bibinfo{volume}{118}},
  \bibinfo{pages}{2977} (\bibinfo{year}{2003}).

\bibitem[{\citenamefont{Bennett}(1976)}]{bennett}
\bibinfo{author}{\bibfnamefont{C.~H.} \bibnamefont{Bennett}},
  \bibinfo{journal}{J.\ Comp.\ Phys.} \textbf{\bibinfo{volume}{22}},
  \bibinfo{pages}{245} (\bibinfo{year}{1976}).

\bibitem[{\citenamefont{Amadei et~al.}(1996)\citenamefont{Amadei, Apol,
  Di{N}ola, and Berendsen}}]{amadei-moments}
\bibinfo{author}{\bibfnamefont{A.}~\bibnamefont{Amadei}},
  \bibinfo{author}{\bibfnamefont{M.~E.~F.} \bibnamefont{Apol}},
  \bibinfo{author}{\bibfnamefont{A.}~\bibnamefont{Di{N}ola}}, \bibnamefont{and}
  \bibinfo{author}{\bibfnamefont{H.~J.~C.} \bibnamefont{Berendsen}},
  \bibinfo{journal}{J.\ Chem.\ Phys.} \textbf{\bibinfo{volume}{104}},
  \bibinfo{pages}{1560} (\bibinfo{year}{1996}).

\bibitem[{\citenamefont{Hummer}(2001)}]{hummer}
\bibinfo{author}{\bibfnamefont{G.}~\bibnamefont{Hummer}}, \bibinfo{journal}{J.\
  Chem.\ Phys.} \textbf{\bibinfo{volume}{114}}, \bibinfo{pages}{7330}
  (\bibinfo{year}{2001}).

\bibitem[{\citenamefont{Pitera and van Gunsteren}(2001)}]{pitera}
\bibinfo{author}{\bibfnamefont{J.~W.} \bibnamefont{Pitera}} \bibnamefont{and}
  \bibinfo{author}{\bibfnamefont{W.~F.} \bibnamefont{van Gunsteren}},
  \bibinfo{journal}{J.\ Phys.\ Chem.\ B} \textbf{\bibinfo{volume}{105}},
  \bibinfo{pages}{11264} (\bibinfo{year}{2001}).

\bibitem[{\citenamefont{Zwanzig}(1954)}]{zwanzig}
\bibinfo{author}{\bibfnamefont{R.~W.} \bibnamefont{Zwanzig}},
  \bibinfo{journal}{J.\ Chem.\ Phys.} \textbf{\bibinfo{volume}{22}},
  \bibinfo{pages}{1420} (\bibinfo{year}{1954}).

\bibitem[{\citenamefont{Hummer and Szabo}(1996)}]{hummer-moments}
\bibinfo{author}{\bibfnamefont{G.}~\bibnamefont{Hummer}} \bibnamefont{and}
  \bibinfo{author}{\bibfnamefont{A.}~\bibnamefont{Szabo}},
  \bibinfo{journal}{J.\ Chem.\ Phys.} \textbf{\bibinfo{volume}{105}},
  \bibinfo{pages}{2004} (\bibinfo{year}{1996}).

\bibitem[{\citenamefont{Hendrix and Jarzynski}(2001)}]{hendrix}
\bibinfo{author}{\bibfnamefont{D.~A.} \bibnamefont{Hendrix}} \bibnamefont{and}
  \bibinfo{author}{\bibfnamefont{C.}~\bibnamefont{Jarzynski}},
  \bibinfo{journal}{J.\ Chem.\ Phys.} \textbf{\bibinfo{volume}{114}},
  \bibinfo{pages}{5974} (\bibinfo{year}{2001}).

\bibitem[{\citenamefont{Crooks}(2000)}]{crooks-pre}
\bibinfo{author}{\bibfnamefont{G.~E.} \bibnamefont{Crooks}},
  \bibinfo{journal}{Phys.\ Rev.\ E} \textbf{\bibinfo{volume}{61}},
  \bibinfo{pages}{2361} (\bibinfo{year}{2000}).

\bibitem[{\citenamefont{Kirkwood}(1935)}]{kirkwood}
\bibinfo{author}{\bibfnamefont{J.~G.} \bibnamefont{Kirkwood}},
  \bibinfo{journal}{J.\ Chem.\ Phys.} \textbf{\bibinfo{volume}{3}},
  \bibinfo{pages}{300} (\bibinfo{year}{1935}).

\bibitem[{\citenamefont{Liu et~al.}(1996)\citenamefont{Liu, Mark, and van
  Gunsteren}}]{liu}
\bibinfo{author}{\bibfnamefont{H.}~\bibnamefont{Liu}},
  \bibinfo{author}{\bibfnamefont{A.~E.} \bibnamefont{Mark}}, \bibnamefont{and}
  \bibinfo{author}{\bibfnamefont{W.~F.} \bibnamefont{van Gunsteren}},
  \bibinfo{journal}{J.\ Phys.\ Chem.} \textbf{\bibinfo{volume}{100}},
  \bibinfo{pages}{9485} (\bibinfo{year}{1996}).

\bibitem[{\citenamefont{Swanson et~al.}(2004)\citenamefont{Swanson, Henchman,
  and Mc{C}ammon}}]{swanson}
\bibinfo{author}{\bibfnamefont{J.~M.~J.} \bibnamefont{Swanson}},
  \bibinfo{author}{\bibfnamefont{R.~H.} \bibnamefont{Henchman}},
  \bibnamefont{and} \bibinfo{author}{\bibfnamefont{J.~A.}
  \bibnamefont{Mc{C}ammon}}, \bibinfo{journal}{Biophys.\ J.}
  \textbf{\bibinfo{volume}{86}}, \bibinfo{pages}{67} (\bibinfo{year}{2004}).

\bibitem[{\citenamefont{Hermans}(1991)}]{hermans}
\bibinfo{author}{\bibfnamefont{J.}~\bibnamefont{Hermans}},
  \bibinfo{journal}{J.\ Phys.\ Chem.} \textbf{\bibinfo{volume}{95}},
  \bibinfo{pages}{9029} (\bibinfo{year}{1991}).

\bibitem[{\citenamefont{Leach}(2001)}]{leach-book}
\bibinfo{author}{\bibfnamefont{A.~L.} \bibnamefont{Leach}},
  \emph{\bibinfo{title}{Molecular Modelling Principles and Applications --
  Second Ed.}} (\bibinfo{publisher}{Prentice Hall}, \bibinfo{address}{Dorset},
  \bibinfo{year}{2001}).

\bibitem[{\citenamefont{Lybrand et~al.}(1985)\citenamefont{Lybrand, Ghosh, and
  Mc{C}ammon}}]{lybrand}
\bibinfo{author}{\bibfnamefont{T.~P.} \bibnamefont{Lybrand}},
  \bibinfo{author}{\bibfnamefont{I.}~\bibnamefont{Ghosh}}, \bibnamefont{and}
  \bibinfo{author}{\bibfnamefont{J.~A.} \bibnamefont{Mc{C}ammon}},
  \bibinfo{journal}{J.\ Am.\ Chem.\ Soc.} \textbf{\bibinfo{volume}{107}},
  \bibinfo{pages}{7793} (\bibinfo{year}{1985}).

\bibitem[{\citenamefont{Frenkel and Smit}(1996)}]{frenkel-book}
\bibinfo{author}{\bibfnamefont{D.}~\bibnamefont{Frenkel}} \bibnamefont{and}
  \bibinfo{author}{\bibfnamefont{B.}~\bibnamefont{Smit}},
  \emph{\bibinfo{title}{Understanding Molecular Simulation}}
  (\bibinfo{publisher}{Academic Press}, \bibinfo{address}{San Diego},
  \bibinfo{year}{1996}).

\bibitem[{\citenamefont{Nanda and Woolf}()}]{nanda}
\bibinfo{author}{\bibfnamefont{H.}~\bibnamefont{Nanda}} \bibnamefont{and}
  \bibinfo{author}{\bibfnamefont{T.~B.} \bibnamefont{Woolf}}, \bibinfo{note}{in
  preparation.}

\bibitem[{tha()}]{thanks}
\bibinfo{note}{Thanks to Hirsh Nanda and Tom Woolf for the raw stearic to
  palmitic acid data, and to David Hendrix and Chris Jarzysnki for the raw
  chemical potential data.}

\bibitem[{\citenamefont{Ponder}(2003)}]{ponder}
\bibinfo{author}{\bibfnamefont{J.~W.} \bibnamefont{Ponder}},
  \bibinfo{journal}{St.\ Louis, MO}  (\bibinfo{year}{2003}).

\bibitem[{\citenamefont{Berendsen et~al.}(1984)\citenamefont{Berendsen, Postma,
  van Gunsteren, Di{N}ola, and Haak}}]{berendsen}
\bibinfo{author}{\bibfnamefont{H.~J.~C.} \bibnamefont{Berendsen}},
  \bibinfo{author}{\bibfnamefont{J.~P.~M.} \bibnamefont{Postma}},
  \bibinfo{author}{\bibfnamefont{W.~F.} \bibnamefont{van Gunsteren}},
  \bibinfo{author}{\bibfnamefont{A.}~\bibnamefont{Di{N}ola}}, \bibnamefont{and}
  \bibinfo{author}{\bibfnamefont{J.~R.} \bibnamefont{Haak}},
  \bibinfo{journal}{J.\ Chem.\ Phys.} \textbf{\bibinfo{volume}{81}},
  \bibinfo{pages}{3684} (\bibinfo{year}{1984}).

\bibitem[{\citenamefont{Andersen}(1983)}]{andersen}
\bibinfo{author}{\bibfnamefont{H.~C.} \bibnamefont{Andersen}},
  \bibinfo{journal}{J.\ Comp.\ Phys.} \textbf{\bibinfo{volume}{52}},
  \bibinfo{pages}{24} (\bibinfo{year}{1983}).

\bibitem[{\citenamefont{Efron and Tibshirani}(1993)}]{bootstrap-book}
\bibinfo{author}{\bibfnamefont{B.}~\bibnamefont{Efron}} \bibnamefont{and}
  \bibinfo{author}{\bibfnamefont{R.~J.} \bibnamefont{Tibshirani}},
  \emph{\bibinfo{title}{An Introduction to the Bootstrap}}
  (\bibinfo{publisher}{Chapman and Hall}, \bibinfo{address}{{New York}},
  \bibinfo{year}{1993}).

\bibitem[{\citenamefont{Politis et~al.}(1999)\citenamefont{Politis, Romano, and
  Wolf}}]{subsample-book}
\bibinfo{author}{\bibfnamefont{D.~N.} \bibnamefont{Politis}},
  \bibinfo{author}{\bibfnamefont{J.~P.} \bibnamefont{Romano}},
  \bibnamefont{and} \bibinfo{author}{\bibfnamefont{M.}~\bibnamefont{Wolf}},
  \emph{\bibinfo{title}{Subsampling}} (\bibinfo{publisher}{Springer-Verlag},
  \bibinfo{address}{{New York}}, \bibinfo{year}{1999}).

\end{thebibliography}

\end{document}